\documentclass{ws-procs9x6}

\usepackage{bm}  % for boldmath
\usepackage{verbatim} % for commenting large sections
\usepackage{scalefnt}
\usepackage{graphicx}
\begin{document}

%\title{Electromagnetic non-contact gears: 
%A prelude to lateral force between corrugated dielectric slabs.}
%Prelude}
\title{ELECTROMAGNETIC NON-CONTACT GEARS: PRELUDE}

\author{Prachi Parashar$^*$ and Kimball A. Milton$^\dagger$}
\address{Oklahoma Center for High Energy Physics
and Homer L. Dodge Department of Physics and Astronomy,
University of Oklahoma, Norman, OK 73019, USA \\
E-mail: $^*$prachi@nhn.ou.edu,
$^\dagger$milton@nhn.ou.edu}
%\homepage{http://www.nhn.ou.edu/%7Emilton}

\author{In\'{e}s Cavero-Pel\'{a}ez}
\address{Theoretical Physics Department, 
Zaragoza University, Zaragoza 50009, Spain\\
E-mail: cavero@unizar.es}

\author{K. V. Shajesh}
\address{Saint Edward's School, Vero Beach, FL 32963, USA\\
E-mail: shajesh@nhn.ou.edu}
%\homepage{http://www.nhn.ou.edu/%7Eshajesh}

%\begin{center}(\today)\end{center}
%\keywords{} 

%--------------------------------------------
\begin{abstract}
We calculate the lateral Lifshitz force between corrugated dielectric slabs 
of finite thickness. Taking the thickness of the plates to infinity 
leads us to the lateral Lifshitz force between corrugated 
dielectric surfaces of infinite extent. Taking the dielectric constant
to infinity leads us to the conductor limit which has been 
evaluated earlier in the literature.
\end{abstract}

%\maketitle
%--------------------------------------------
%\tableofcontents
%--------------------------------------------

\section{Introduction}

In past decade significant attention has been given to evaluation of 
the lateral force between corrugated surfaces 
(for example see Ref.~\refcite{Emig:2003,Chen:2002zz,Lambrecht:2008zz,
CaveroPelaez:2008tj,CaveroPelaez:2008tk} and references there-in).
In an earlier work we calculated the contribution of the next-to-leading 
order to the lateral Casimir force between two corrugated semi-transparent 
$\delta$-function plates interacting with 
a scalar field~\cite{CaveroPelaez:2008tj}, 
and the leading order contribution for the case of two concentric 
semi-transparent corrugated cylinders\cite{CaveroPelaez:2008tk} 
using the multiple scattering formalism
(see Ref.~\refcite{Milton:2007wz,Milton:2008st} and references there-in).
We observed that including the next-to-leading order contribution 
significantly reduced the deviation from the exact result in 
the case of weak coupling. 
Comparison with experiments requires the analogous calculation for
the electromagnetic case.
Here we present preliminary results of our ongoing work on
the evaluation of the lateral Lifshitz force 
between two corrugated dielectric (non-magnetic) slabs of finite thickness
interacting through the electromagnetic field (see Fig. \ref{corru}). 

From the general result it is easy to take various limiting cases.
Taking the thickness of the dielectric slabs to infinity leads us 
to the lateral Lifshitz force between dielectric slabs of infinite extent. 
The lateral Casimir force between corrugated conductors
was evaluated by Emig et al\cite{Emig:2003}. In our situation this 
is achieved by taking the dieletric constants
$\varepsilon_i \rightarrow \infty$. Our results 
agree with the results in  Emig et al\cite{Emig:2003}.
Taking the thin-plate approximation based on the plasma model we 
have calculated the lateral force between corrugated plasma sheets.
Our goal is to extend these results to next-to-leading order.
Most of these will appear in a forthcoming paper.

%--------------------------------------------
\section{Interaction energy}

\begin{figure}
\begin{center}
\includegraphics[width=80mm,scale=1.0]{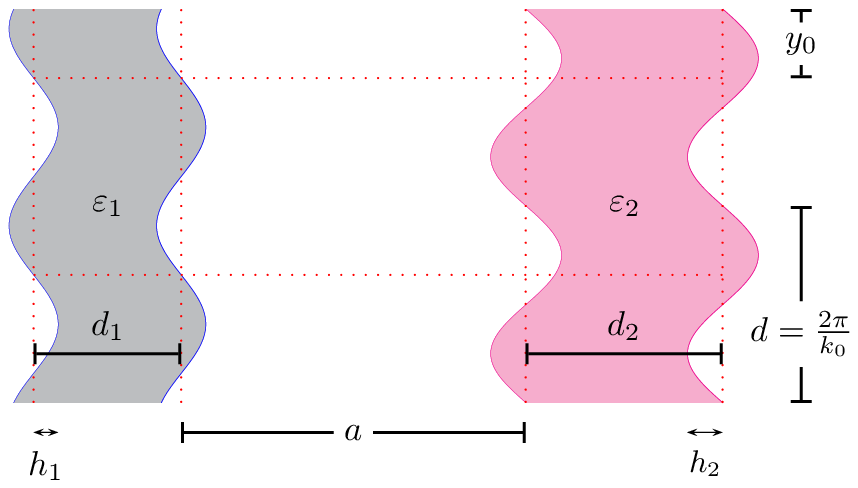}
\caption{Parallel dielectric slabs with sinusoidal corrugations.}
\label{corru}
\end{center}
\end{figure}

We consider two dielectric slabs of infinite extent in $x$-$y$ 
plane, which have corrugations in $y$-direction,
as described in Fig.\,\ref{corru}. We describe the dielectric slabs 
by the potentials
\begin{equation}
V_i(z,y) = (\varepsilon_i-1)\,
        \left[\theta (z - a_i - h_i(y))- \theta(z - b_i - h_i(y))\right],
\label{pot-i}
\end{equation}
where $i=1,2$, designates the individual dielectric slabs. 
$\theta(z)$ is the Heaviside theta function 
defined to equal 1 for $z>0$, and 0 when $z<0$.
$h_i(y)$  describes the corrugations on the surface of the slabs. 
We define the thickness of the individual slabs as $d_i=b_i-a_i$, 
such that $a = a_2 - b_1 > 0$ represents the distance between the slabs.
The permittivities of the slabs are represented by $\varepsilon_i$.
%with $i=a,b$ for clarity in notation.
%%%%%%%%%%%%%%%%%%%%%%%%%%%%%%%%%%%%%%%
%%%%%%%%%%%%%%%%%%%%%%%%%%%%%%%%%%%%%%%%%%

Using the multiple scattering formalism for the case of the electromagnetic 
field~\cite{Milton:2008vr,Milton:2008bb} 
based on Schwinger's Green's dyadic formalism~\cite{Schwinger:1977pa}
and following the formalism described in Gears-I~\cite{CaveroPelaez:2008tj} 
we can obtain the contribution to the interaction energy between the 
two slabs in leading order in the corrugation amplitudes to be
\begin{equation}
E_{12}^{(2)}=\frac{i}{2}\,\int \frac{d\omega}{2\pi}
\,\text{Tr} \left[\bm{\Gamma}^{(0)} \Delta V_1^{(1)}
\cdot \bm{\Gamma}^{(0)}\,\Delta V_2^{(1)} \right],
\label{intE2}
\end{equation}
where $\Delta V_i^{(1)}$ are the leading order contributions in the 
potentials due to the presence of corrugations. In particular,
we have
\begin{equation}
\Delta V_i^{(1)}(z,y) 
= -h_i(y) \left(\varepsilon_i-1\right)
        \left[\delta (z - a_i)-\delta(z - b_i)\right].
\label{pot-i1}
\end{equation}
Note that $V_i^{(0)}$ describes the potential for the case when 
the corrugations are absent and represent the background in the 
formalism.
$\bm{\Gamma}^{(0)}=\bm{\Gamma}^{(0)}(\bm{x},\bm{x}^\prime;\omega)$ 
is the Green's dyadic in the presence of background potential $V_i^{(0)}$ 
and satisfies
\begin{equation}
\left[-\frac{1}{\omega^2}\bm{\nabla}\times\bm{\nabla}\times 
\;+\; \bm{1} + V_1^{(0)} + V_2^{(0)} \right]
\cdot \bm{\Gamma}^{(0)} =-\bm{1}.
\label{G(0)eq}
\end{equation}
The corresponding reduced Green's dyadic 
$\,\bm{\gamma}^{(0)}(z,z^\prime;k_x,k_y,\omega)$ is defined by 
Fourier transforming in the transverse variables as
\begin{equation}
\bm{\Gamma}^{(0)}(\bm{x},\bm{x}^\prime;\omega)
= \int\frac{dk_x}{2\pi}\frac{dk_y}{2\pi}
\,e^{ik_x(x-x^\prime)} e^{ik_y(y-y^\prime)}
\,\bm{\gamma}^{(0)}(z,z^\prime;k_x,k_y,\omega).
\end{equation} 
Using the fact that our system is translationally invariant 
in the $x$-direction, we can write
\begin{equation}
\frac{E_{12}^{(2)}}{L_x}
= \int_{-\infty}^{\infty} \frac{dk_{y}}{2\pi}
  \int_{-\infty}^{\infty} \frac{dk_{y}^\prime}{2\pi}
\,\tilde{h}_1(k_{y}-k_{y}^\prime) \,\tilde{h}_2(k_{y}^\prime-k_{y}) 
\,L^{(2)}(k_{y},k_{y}^\prime),
\label{DE12=L2}
\end{equation}
where $L_x$ is the length in the $x$-direction and 
$\tilde{h}_i(k_y)$ are the Fourier transforms of the 
functions $h_i(y)$ describing the corrugations. 
The kernel $L^{(2)}(k_{y},k_{y}^\prime)$ is given by
\begin{eqnarray}
L^{(2)}(k_{y},k_{y}^\prime) 
= -\frac{1}{2}\int \frac{d\zeta}{2 \pi} 
\int \frac{d k_x}{2 \pi}\, I^{(2)}(k_x,\zeta,k_{y},k_{y}^\prime),
\label{L2}
\end{eqnarray}
where
\begin{eqnarray}
I^{(2)}(k_x,\zeta,k_{y},k_{y}^\prime) 
&=& \left(\varepsilon_a-1\right) \left(\varepsilon_b-1\right) 
\hspace{40mm} \nonumber\\ &&
\times \Big[ \,\bm{\gamma}^{(0)}(a_2,a_1;k_x,k_{y},\omega)
\cdot \bm{\gamma}^{(0)}(a_1,a_2;k_x,k_{y}^\prime,\omega) 
\nonumber \\ && \hspace{3mm}
-\,\bm{\gamma}^{(0)}(b_2,a_1;k_x,k_{y},\omega)
\cdot \bm{\gamma}^{(0)}(a_1,b_2;k_x,k_{y}^\prime,\omega) 
\nonumber \\ && \hspace{3mm}
-\,\bm{\gamma}^{(0)}(a_2,b_1;k_x,k_{y},\omega)
\cdot \bm{\gamma}^{(0)}(b_1,a_2;k_x,k_{y}^\prime,\omega) 
\nonumber \\ && \hspace{3mm}
+\,\bm{\gamma}^{(0)}(b_2,b_1;k_x,k_{y},\omega)
\cdot \bm{\gamma}^{(0)}(b_1,b_2;k_x,k_{y}^\prime,\omega) \Big],
\label{I2}
\end{eqnarray}
where the reduced Green's dyadics are evaluated after 
solving Eq.\,(\ref{G(0)eq}). 
We note that
$\bm{\gamma}^{(0)}(z,z^\prime;k_x,k_{y},\omega)
={\bm{\gamma}^{(0)}}^\dagger(z^\prime,z;k_x,k_{y},\omega)$.
Our task reduces to evaluating the reduced Green's dyadic 
in the presence of the background. 
The details of this evaluation will be described in the 
forthcoming paper.

%%%%%%%%%%%%%%%%%%%%%%%%%%%%%%%%%%%%%%%%%%%%%%%%%%%%%%
\subsection{Evaluation of the reduced Green's dyadic}

The Green's dyadic satisfies Eq.\,(\ref{G(0)eq}) whose solution
can be determined by following the procedure decribed 
in Schwinger et al\cite{Schwinger:1977pa}. 
The expression for the reduced Green's dyadic
%$\mathbf{g}^{(0)}(z,z^{\prime};k,0,\zeta) =$
\begin{equation}
\bm{\gamma}^{(0)}(z,z^{\prime};k,0,\zeta) =
\left[ \begin{array}{ccc}
\frac{1}{\varepsilon (z)} \frac{\partial}{\partial z}
\frac{1}{\varepsilon (z^\prime)} \frac{\partial}{\partial z^\prime}
\,g^H%(z,z^\prime)
& 0 & 
\frac{ik}{\varepsilon (z^\prime)}
\frac{1}{\varepsilon (z)} \frac{\partial}{\partial z}
\,g^H%(z,z^\prime) 
\\[3mm]
0 & -\zeta^2 \,g^E%(z,z^\prime) 
& 0 \\[3mm]
-\frac{ik}{\varepsilon (z)}
\frac{1}{\varepsilon (z^\prime)} \frac{\partial}{\partial z^\prime}
\,g^H%(z,z^\prime)
& 0 & \frac{k^2}{\varepsilon (z) \varepsilon (z^\prime)}
\,g^H%(z,z^\prime)
\end{array} \right]
\label{spG0}
\end{equation}
is given in terms of the electric and magnetic Green's 
%----footnote
functions\footnote{Here we use the notation in 
Schwinger et al\cite{Schwinger:1977pa}
which was reversed in many of Milton's publications,
for example in Milton's book\cite{Milton:2001-book}.}
%----end of footnote
$g^E(z,z^\prime)$ and $g^H(z,z^\prime)$, which satisfy the 
following differential equations: 
\begin{eqnarray}
- \left[ \frac{\partial^2}{\partial z^2} - k^2 - \zeta^2 \varepsilon (z)
\right] \,g^E(z,z^\prime) = \delta (z-z^\prime),
\\- \left[ 
\frac{\partial}{\partial z} 
\frac{1}{\varepsilon (z)} \frac{\partial}{\partial z}
- \frac{k^2}{\varepsilon (z)} - \zeta^2 
\right] \,g^H(z,z^\prime) = \delta (z-z^\prime).
\end{eqnarray}
We have used the definitions $k^2 = k_x^2 + k_y^2$
and $\varepsilon (z) = 1+V_1^{(0)}(z)+V_2^{(0)}(z)$.

The reduced Green's dyadic for arbitrary $k_y$ is generated by the rotation
\begin{equation}
\bm{\gamma}^{(0)}(z,z^{\prime};k_x,k_y,\zeta)
= \mathbf{R} \cdot \bm{\gamma}^{(0)}(z,z^{\prime};k,0,\zeta) \cdot \mathbf{R}^T,
\end{equation} 
where
\begin{equation}
\mathbf{R} = \frac{1}{k}
\left( \begin{array}{ccc}
k_x & -k_y & 0 \\
k_y & k_x & 0 \\
0 & 0 & k 
\end{array} \right).
\end{equation}
We have dropped delta functions in Eq.\,(\ref{spG0}) because they 
are evaluated at different points and thus do not contribute.
We shall not present explicit solutions to the electric and magnetic 
Green's functions here which will be presented in our forthcoming paper.

%%%%%%%%%%%%%%%%%%%%%%%%%%%%%%%%%%%%%%%%%%%%%%%%%%%%%%%%%%%%%%%
\subsection{Interaction energy for corrugated dielectric slabs}

Using the solutions to the electric and magnetic Green's function
in Eq.\,(\ref{spG0}) we can evaluate $I^{(2)}(k_x,\zeta,k_{y},k_{y}^\prime)$ 
in Eq.\,\eqref{I2} as
\begin{align}
%I^{(2)}(k_x,\zeta,k_{y},k_{y}^\prime) &= 
-\,&\frac{1}{k^2} \frac{1}{{k^\prime}^2}
\frac{1}{2\kappa} \frac{1}{2\kappa^\prime} 
\Bigg[
\frac{1}{\Delta} \frac{1}{\Delta^\prime}
      M(-\alpha_1,-\alpha_1^\prime) M(-\alpha_2,-\alpha_2^\prime)
(k_x^2+k_yk_y^\prime)^2 \zeta^4
\hspace{30mm} \nonumber \\
&+   \frac{1}{\Delta} \frac{1}{\bar{\Delta}^\prime}
      M(-\alpha_1,\bar{\alpha}_1^\prime) M(-\alpha_2,\bar{\alpha}_2^\prime)
k_x^2(k_y-k_y^\prime)^2\zeta^2{\kappa^\prime}^2
\hspace{30mm} \nonumber \\
&+   \frac{1}{\bar{\Delta}} \frac{1}{\Delta^\prime}
      M(\bar{\alpha}_1,-\alpha_1^\prime) M(\bar{\alpha}_2,-\alpha_2^\prime)
k_x^2(k_y-k_y^\prime)^2\zeta^2\kappa^2
\hspace{30mm} \nonumber \\
&+ \frac{1}{\bar{\Delta}} \frac{1}{\bar{\Delta}^\prime}
\left\{ M(\bar{\alpha}_1,\bar{\alpha}_1^\prime) 
	(k_x^2+k_yk_y^\prime)\kappa\kappa^\prime
        + M(-\bar{\alpha}_1,-\bar{\alpha}_1^\prime) 
          k^2 {k^\prime}^2 \frac{1}{\varepsilon_1} \right\}
\hspace{30mm} \nonumber \\
&\hspace{10mm}\times
\left\{ M(\bar{\alpha}_2,\bar{\alpha}_2^\prime) 
        (k_x^2+k_yk_y^\prime)\kappa\kappa^\prime
        + M(-\bar{\alpha}_2,-\bar{\alpha}_2^\prime)
          k^2 {k^\prime}^2 \frac{1}{\varepsilon_2} \right\}
\Bigg],
\label{trGG}
\end{align} 
where
\begin{eqnarray}
\Delta &=& 
\big[ (1-\alpha_1^2\,e^{-2\kappa_1 d_1})
                  (1-\alpha_2^2\,e^{-2\kappa_2 d_2})\,e^{\kappa a}
\nonumber \\ && \hspace{5mm}
-\alpha_1\alpha_2 (1-e^{-2\kappa_1 d_1})
                  (1-e^{-2\kappa_2 d_2})\,e^{-\kappa a} \big],
\\
M(\alpha_i,\alpha_i^\prime)
&=& (\varepsilon_i-1) \big[ (1-\alpha_i^2)\,e^{-\kappa_i d_i}
                  (1-{\alpha_i^\prime}^2)\,e^{-\kappa_i^\prime d_i}
\nonumber \\ && \hspace{5mm}
-(1+\alpha_i)(1-\alpha_i\,e^{-2\kappa_i d_i})
 (1+\alpha_i^\prime)(1-\alpha_i^\prime \,e^{-2\kappa_i^\prime d_i}) \big],
\end{eqnarray}
where $\kappa_i^2 = k^2 + \zeta^2 \varepsilon_i$,
$\bar{\kappa}_i=\kappa_i/\varepsilon_i$, and
$\alpha_i = (\kappa_i -\kappa)/(\kappa_i +\kappa)$.
Quantities with primes are obtained by replacing 
$k_y\rightarrow k_y^\prime$ everywhere,
and quantities with bars are obtained by replacing $\kappa_i$ with 
$\bar{\kappa}_i$ except in the exponentials.

%%%%%%%%%%%%%%%%%%%%%%%%%%%%%%%%%%%%%%%%%%%%%%%%%
\subsection{Conductor limit}

\begin{figure}
\begin{center}
\includegraphics[width=80mm,scale=1.0]{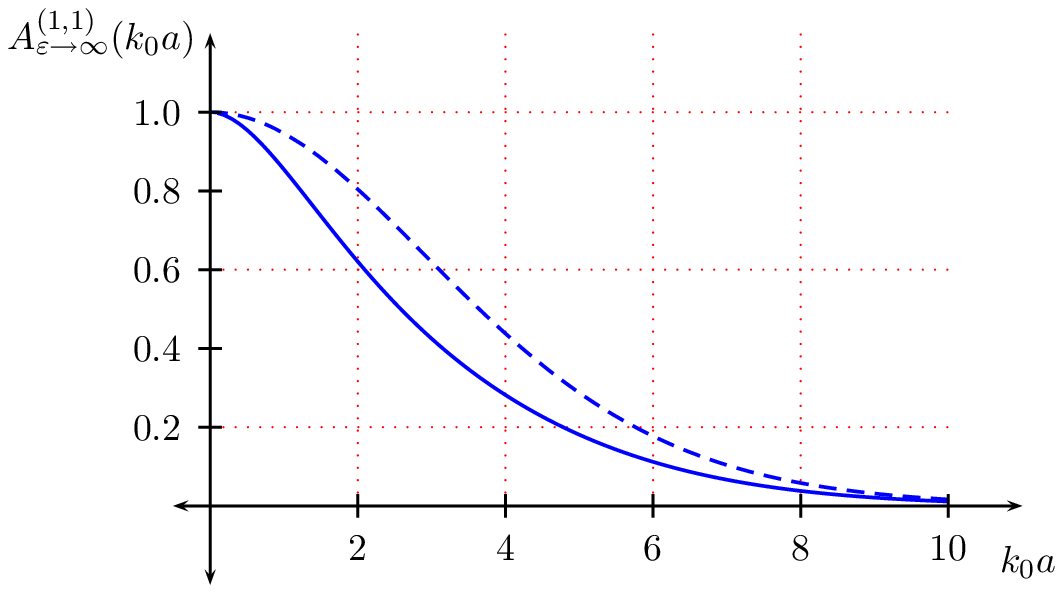}
\caption{Plot of $A^{(1,1)}_{\varepsilon\rightarrow\infty}(k_0a)$
versus $k_0a$. The dotted curve represents 2 times the Dirichlet case.}
\label{Aem11-cond-versus-t0}
\end{center}
\end{figure}

In the conductor limit ($\varepsilon_i\rightarrow\infty$) 
the above expression takes the form
\begin{equation}
I^{(2)}_{\varepsilon \rightarrow\infty}
(\kappa,\kappa^\prime,k_y-k_y^\prime) =
-\frac{\kappa}{\sinh \kappa a}
\frac{\kappa^\prime}{\sinh \kappa^\prime a}
\left[ 1 +
\frac{\{\kappa^2 + {\kappa^\prime}^2 - (k_y-k_y^\prime)^2\}^2}
     {4\,\kappa^2 {\kappa^\prime}^2}
\right].
\end{equation}
For the case of sinusoidal corrugations described by
$h_1(y)=h_1\sin [k_0(y+y_0)]$ and
$h_2(y)=h_2\sin [k_0y]$ the lateral force can be evaluated to be
\begin{equation}
F^{(2)}_{\varepsilon\rightarrow\infty}
= 2k_0a\, \sin (k_0y_0) \left|F^{(0)}_\text{Cas}\right|
\frac{h_1}{a} \frac{h_2}{a}
A^{(1,1)}_{\varepsilon\rightarrow\infty} (k_0a),
\end{equation}
where
\begin{equation}
A^{(1,1)}_{\varepsilon\rightarrow\infty} (t_0)
= \frac{15}{\pi^4}\int_{-\infty}^\infty dt \int_0^\infty \bar{s}d\bar{s}
\frac{s}{\sinh s} \frac{s_+}{\sinh s_+}
\left[ \frac{1}{2} + 
\frac{(s^2 + s_+^2 - t_0^2)^2}{8\,s^2s_+^2} \right],
\label{A-cond-2int}
\end{equation}
where $s^2=\bar{s}^2+t^2$ and $s_+^2=\bar{s}^2+(t+t_0)^2$.
The first term in  Eq.\,\eqref{A-cond-2int} corresponds to the 
Dirichlet scalar case~\cite{CaveroPelaez:2008tj}, which here corresponds 
to the E mode (referred to in Ref.~\refcite{Emig:2003} as TM mode).
We note that $A^{(1,1)}_{\varepsilon\rightarrow\infty} (0)=1$.
See Fig.~\ref{Aem11-cond-versus-t0} for the plot of
$A^{(1,1)}_{\varepsilon\rightarrow\infty}(k_0a)$ versus $k_0a$.
We observe that only in the PFA limit is the electromagnetic 
contribution twice that of the Dirichlet case, and in general 
the electromagnetic case is less than twice that of the Dirichlet case.
%%%%%%%%%%%%%%%%%%%%%%%%%%%%%%%%%%%%%%%
%%%%%%%%%%%%%%%%%%%%%%%%%%%%%%%%%%%%%%%%%%

Since the above expression involves a convolution of two functions
we can evaluate one of the integrals to get
\begin{eqnarray}
A^{(1,1)}_{\varepsilon\rightarrow\infty} (t_0)
&=& \frac{15}{4} \int_0^\infty du\, \frac{\sin (2t_0u/\pi)}{(2t_0u/\pi)}
\Bigg[ \frac{\sinh^2u}{\cosh^6u} \left( \frac{7}{2} - \sinh^2u \right)
\nonumber \\ &&
-\frac{1}{2} \left(\frac{2t_0}{\pi}\right)^2 \frac{\sinh^2 u}{\cosh^4 u} 
+\frac{1}{16} \left(\frac{2t_0}{\pi}\right)^4 \frac{\sinh^2 u}{\cosh^2 u}
\Bigg]
\label{A-cond-1int},
\end{eqnarray}
which reproduces the result in Emig et al~\cite{Emig:2003} 
apart from an overall factor of 2,
which presumably is a transcription error.
%Presumably the lower limits in the integrals in Eq. (48) in 
%Emig et al\cite{Emig:2003} should be $0$ instead of negative infinity.
%This is necessary to have the correct PFA limit ans has been stated 
%correctly after their Eq.\,(48). 
Even though Eq.~\eqref{A-cond-1int} involves only a single integral 
it turns out that the double integral representation in Eq.~\eqref{A-cond-2int}
is more useful for numerical evaluation because of the oscillatory 
nature of the function $\sin x/x$ in the former. 
%---------------------------------------------------------------------
\section{Conclusion}
We have evaluated leading order contribution to the lateral Lifshitz 
force between two corrugated dielectric slabs. Taking the dielectric constants 
of the two bodies to infinity gives the lateral Casimir force between 
corrugated conductors. We shall extend these results to next-to-leading 
order contribution for a better comparison with experiments in future 
publication as well as include various other limiting cases, which 
can be readily obtained from Eq.\,~\eqref{trGG}.
%------------------------------------------------------------------------
\section*{Acknowledgements}
We thank the US Department of Energy for partial support of this work. 
We extend our appreciation to Jef Wagner, Elom Abalo and Nima Pourtolami
for useful comments throughout the work.

%---------------------------------------------------------------
%---------------------------------------------------------------

\end{document}